\documentclass[a4paper,10pt]{article}
\usepackage[utf8x]{inputenc}
\usepackage{graphicx}
\usepackage{amsmath}
\usepackage{amsfonts}
\usepackage{amssymb}

\title{}
\author{}
\newcommand{\msinglet}{m_{S}}
\newcommand{\mhiggs}{m_h}
\newcommand{\oh}{\Omega h^2}
\newcommand{\gev}{\mathrm{GeV}}
\newcommand{\tev}{\mathrm{TeV}}
\newcommand{\sv}{\langle\sigma v\rangle}

\begin{document}
\title{\huge The singlet scalar as FIMP dark matter}
\author{\large Carlos E. Yaguna\footnote{carlos.yaguna@physik.uni-wuerzburg.de}\\ \textit{\normalsize Institute f\"ur Theoretische Physik und Astrophysik,}\\ \textit{ \normalsize Universit\"at W\"urzburg, 97074 W\"urzburg, Germany}}
\date{}
\maketitle

\begin{abstract}
The singlet scalar model is a minimal extension of the Standard Model that can explain the dark matter. We point out that in this model the dark matter constraint can be satisfied
 not only in the already considered WIMP regime but also, for much smaller couplings, in the Feebly Interacting Massive Particle (FIMP) regime. In it, dark matter particles are slowly produced in the early Universe but are never abundant enough to reach thermal equilibrium or annihilate among themselves.  This alternative framework is as simple and predictive as the WIMP scenario but it gives rise to a completely different dark matter phenomenology.  After reviewing the calculation of the dark matter relic density in the FIMP regime, we study in detail the evolution of the dark matter abundance in the early Universe and the predicted relic density as a function of the parameters of the model.  A new dark matter compatible region of the singlet  model is identified, featuring couplings of order $10^{-11}$ to $10^{-12}$ for singlet masses in the $\gev$ to $\tev$ range. As a consequence, no signals at direct or indirect detection experiments are expected. The relevance of this new viable region for the correct interpretation of recent experimental bounds is emphasized.
\end{abstract}
\section{Introduction}
Weakly Interacting Massive Particles (WIMPs) constitute the most studied class of dark matter candidates. They appear in several extensions of the Standard Model and have the virtue of acquiring, via thermal freeze out  in the early Universe, the right relic density to explain the dark matter. Among them, we can mention the neutralino in supersymmetric models \cite{Griest:2000kj}, the lightest Kaluza-Klein particle in Universal Extra-dimensional models \cite{Hooper:2007qk}, as well as the singlet \cite{McDonald:1993ex,Burgess:2000yq} and the doublet \cite{Barbieri:2006dq,LopezHonorez:2006gr,Honorez:2010re} scalar extensions of the Standard Model. At present, in spite of its popularity, the WIMP framework for dark matter has to be regarded as an interesting hypothesis to be tested in current and future experiments. WIMPs  typically give rise to a number of signatures in accelerator searches  and in direct and indirect dark matter detection experiments.  In fact, it has been recently claimed that the moment of truth for WIMPs has finally arrived \cite{Bertone:2010at}; if they  exist they will be discovered in the next $5$ to $10$ years. So far, however, there is no experimental evidence that dark matter is actually composed of WIMPs.

Moreover, viable and well-motivated alternatives to the WIMP framework do exist \cite{Choi:2005vq,Feng:2003xh,Hall:2009bx}. In \cite{Hall:2009bx},  a new calculable mechanism of dark matter production, dubbed \emph{freeze in}, was introduced. It involves a Feebly Interacting Massive Particle (FIMP) which never attains thermal equilibrium in the early Universe. FIMPs are slowly produced by collisions or decays of particles in the thermal plasma and, in contrast to WIMPs, are never abundant enough to annihilate among themselves. Thus, the abundance of FIMPs freezes in with a yield that \emph{increases} with the interaction strength between them and the thermal bath. FIMPs, in addition, give rise to   completely different signatures in colliders and dark matter experiments \cite{Hall:2009bx}. Due to its simplicity and predictivity, the \emph{freeze in} of FIMPs is a strong contender to the WIMP framework for dark matter.

The singlet scalar model \cite{McDonald:1993ex,Burgess:2000yq} is a minimal extension of the Standard Model that can account for the dark matter. It contains just one additional field, an scalar that is singlet under the Standard Model gauge group and odd under a new $Z_2$ symmetry, and two new parameters.  The phenomenology of this model has been extensively studied in a number of previous works (see e.g. \cite{Yaguna:2008hd,Goudelis:2009zz}). All of them, however, have assumed the WIMP solution to the dark matter constraint. In this paper, we point out that the singlet model can also satisfy the relic density constraint in the FIMP regime, and we analyze in detail the new viable region corresponding to that solution. After briefly introducing the model, we review the calculation of the relic density in the FIMP regime of the singlet model. Then, we study the evolution of the dark matter abundance and the predicted relic density as a function of the parameters of the model. The new viable region corresponding to FIMP dark matter is identified --it features couplings between $10^{-11}$ and $10^{-12}$ for singlet masses in the $\gev$ to $\tev$ range. Finally, we discuss the implications of these results for the correct interpretation of recent experimental bounds.   

\section{The singlet scalar model of dark matter}
The singlet scalar model \cite{McDonald:1993ex,Burgess:2000yq}  is one of the minimal extensions of the Standard Model that can explain the dark matter. It contains an additional field, $S$, that is singlet under the SM gauge group and odd under a new $Z_2$ symmetry that guarantees its stability. The Lagrangian that describes this model is
\begin{equation}
\mathcal{L}= \mathcal{L}_{SM}+\frac 12 \partial_\mu S\partial^\mu S-\frac{m_0^2}{2}S^2-\frac{\lambda_S}{4}S^4-\lambda S^2 H^\dagger H\,,
\label{eq:la}
\end{equation}
where $\mathcal{L}_{SM}$ denotes the Standard Model Lagrangian and $H$ is the higgs doublet. The above is the most general renormalizable Lagrangian that is compatible with  the assumed symmetries and field content of the model. The singlet scalar model introduces, therefore, only two new parameters\footnote{$\lambda_S$ is essentially irrelevant.}: the singlet mass ($m_S=\sqrt{m_0^2+\lambda v^2}$) and the coupling to the higgs boson, $\lambda$. In addition, the higgs mass, a SM parameter, also affects the phenomenology of the model.

In the early Universe, singlets are pair-produced as the particles in the thermal plasma scatter off each other. The dominant production processes are  $s$-channel higgs boson mediated diagrams originating in   a variety of initial states: $f\bar f$, $W^+W^-$, $Z^0Z^0$, and $hh$\footnote{The corresponding annihilation processes are obtained by exchanging initial and final states.}. Likewise, they can also be produced from the initial state $hh$ either directly or through singlet exchange. As a general rule, it is the initial state $W^+W^-$ that tends to dominate the total production rate of dark matter in this model.

The main advantage of the singlet model with respect to other models of dark matter is its simplicity, which allows one to make concrete predictions about dark matter observables. In fact, by imposing the relic density constraint, $\Omega_S=\Omega_{dm}$, one can eliminate $\lambda$ and compute the expected rates for direct and indirect detection experiments as a function of the singlet mass only (for a given higgs mass). This program has indeed been implemented in a number of recent works --see e.g. \cite{Yaguna:2008hd,Goudelis:2009zz}. In them, it was demonstrated that the singlet model can be probed in current direct and indirect detection experiments.  

All previous works on the singlet scalar model have assumed the WIMP solution to the dark matter constraint. It has recently been realized, however, that an alternative mechanism of dark matter production, the freeze in of FIMPs, which is as simple and predictive as the freeze out of WIMPs, can also explain naturally the observed value of the dark matter density. The singlet scalar model has the particularity that both solutions to the dark matter constraint, WIMP and FIMP, can be realized --although in very different  regions of the parameter space. In the following, we will focus on the FIMP regime of the singlet scalar model.
\section{The relic density in the FIMP regime}
It is easy to convince oneself that not only in the singlet model but in all models where the interactions of the dark matter particle are determined by a free parameter $\lambda$, the dark matter constraint can be satisfied not only in the usual WIMP regime, where $\lambda$ typically lies between $10^{-1}$ and $10^{-3}$, but also for much smaller couplings --in the FIMP regime. Imagine, for instance, that for $m=m_{dm}$ and  $\lambda=\lambda_{wimp}$ the relic density constraint is satisfied in the WIMP regime, and let us now analyze how the relic density changes as we go from $\lambda=\lambda_{wimp}$ to $\lambda=0$. Initially, the relic density will increase, for a smaller $\lambda$ implies a smaller annihilation rate and an earlier freeze-out. That is the well-known behavior of WIMP particles. It is clear, however, that such behavior cannot continue for long because the relic density should go to zero rather than to infinity as $\lambda\to 0$. In fact, if $\lambda=0$ the dark matter particle does not interact with other particles and  cannot be thermally produced in the early Universe\footnote{Throughout this paper we do not consider non-thermal production mechanisms.}. Hence, at some point the relic density  starts to decrease as $\lambda$ becomes smaller and it will continue to do so until it vanishes for $\lambda=0$. Since the dark matter relic density is a continuous function of $\lambda$,  a smaller value of $\lambda$, $\lambda_{fimp}$, must exist such that the relic density is also consistent with the observed dark matter density. This alternative solution to the dark matter constraint in the singlet model had not been considered before.

In the FIMP regime, dark matter particles do not reach equilibrium in the early Universe and are never abundant enough to annihilate among themselves. As a result, the dark matter abundance, $Y=n/s$, is not determined by the same processes as in the WIMP framework. If dark matter particles are pair produced, as it happens in the singlet model,  $Y$ satisfies instead the Boltzmann equation
\begin{equation}
\frac{dY}{dT}=\sqrt{\frac{\pi g_*(T)}{45}}M_p\langle\sigma v\rangle Y_{eq}(T)^2
\label{eq:boltzmann}
\end{equation}
with  the boundary condition $Y(T>>m_{dm})=0$. In the above equation $M_p$ is the Planck mass and  $\langle\sigma v\rangle$ is the usual thermally averaged production (or annihilation) cross section. It is through this quantity that the dependence on the particle physics model enters into the evolution of the dark matter abundance. In the singlet scalar model we have that $\sv\propto \lambda^2$. The main difference between the WIMP and the FIMP regimes of the singlet model is the value of $\lambda$ --or $\langle\sigma v\rangle$--, which, being much smaller for FIMPs, prevents them from ever reaching thermal equilibrium: $Y\ll Y_{eq}$. For that reason, the annihilating term proportional to $Y^2$ does not contribute to the right-hand side of equation (\ref{eq:boltzmann}), and it  cannot be assumed, as is the case for  WIMPs, that the dark matter particle was in equilibrium, $Y(T)=Y_{eq}(T)$, for $T\sim m_{dm}$.

Using equation (\ref{eq:boltzmann}) we can anticipate the generic behavior of $Y(T)$ in a large class of models including the singlet scalar. Since the right-hand side is greater or equal to zero, the abundance either increases or remains constant but never decreases,  in agreement with the expectation that dark matter annihilations, which would reduce the abundance, play no role in this regime.  At high temperatures, $Y_{eq}$ is constant and since all particles are relativistic,  we typically have that $\sv\propto 1/T^2$. In that region then, $Y(T)\propto 1/T$.  At low temperatures, $T<<m_{dm}$, the particles in the thermal plasma no longer have enough energy to produce dark matter particles in their scatterings, $\sv\to 0$, so the right-hand side goes to zero and $Y(T)$ remains constant. Thus, the dark matter abundance initially increases as the Universe cools down but at a certain point it reaches the \emph{freeze in} temperature, below which the abundance no longer changes. In the next section, we will numerically confirm this simple behavior of the dark matter abundance for  low and high temperatures.

Integrating equation (\ref{eq:boltzmann}) from $T_i\gg m_{dm}$ to $T_f\ll m_{dm}$ leads to the  current abundance of dark matter, $Y(T_0)$. From it, the dark matter relic density can be calculated in the usual way:
\begin{equation}
\oh=2.742\times 10^8 \frac{m_{dm}}{\gev} Y(T_0). 
\label{eq:rd}
\end{equation}
 
Notice that the relic density of dark matter in this regime is also determined by thermal processes. Besides the given particle physics model, both regimes (WIMP and FIMP) assume only the Standard Cosmological Model --e.g. a radiation dominated Universe during  dark matter production. Regarding dark matter, henceforth, they are equally predictive and should be treated on the same footing. In the singlet scalar model, there is no objective reason to prefer the WIMP solution to the dark matter constraint over the FIMP one. 
 
Finally, we would like to mention that in spite of the similarity between the equations that determine the evolution of the dark matter abundance in the WIMP and FIMP regimes, it is usually not possible to compute the FIMP relic density using the numerical codes written for WIMPs. MicrOMEGAs\cite{Belanger:2010gh}, in particular,  assumes the WIMP paradigm (e.g. $Y(T\sim m)=Y_{eq}$), giving wrong results if applied directly to the FIMP framework. In our numerical analysis, we have used micrOMEGAs but only to calculate $\sv$ and have solved equation (\ref{eq:boltzmann}) for ourselves. In the next section we present the main results of this paper. 
\section{Results}
\begin{figure}[tb]
\begin{center}
\includegraphics[scale=0.45]{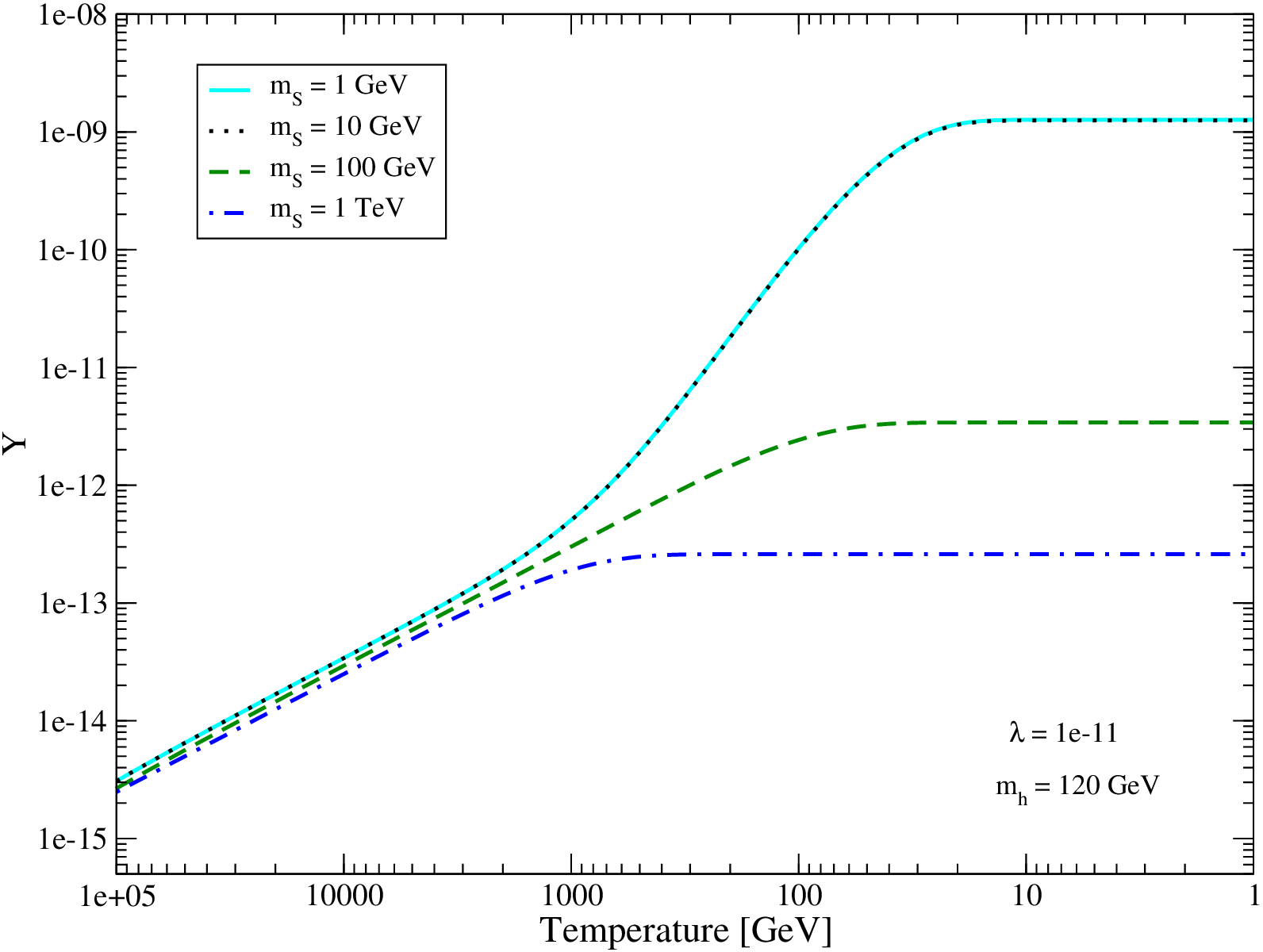}
\caption{\small \it The dark matter abundance, $Y$, as a function of the temperature for different values of $\msinglet$. In this figure $\mhiggs$ was set to $120~\gev$ and $\lambda$ to $10^{-11}$. 
\label{fig:l-11}}
\end{center}
\end{figure}

In this section, we compute the evolution of the dark matter abundance and the predicted relic density as  a function of the parameters of the singlet model: $\lambda$, $\msinglet$ and $\mhiggs$. We also find the new viable region corresponding to the FIMP solution to the dark matter constraint.

The evolution of the dark matter abundance as a function of the temperature is shown in figure \ref{fig:l-11} for different singlet masses between $1~\gev$ and $1~\tev$. In it, $\lambda$ and $\mhiggs$ were fixed respectively to $10^{-11}$ and $120~\gev$. The general behavior of $Y$ is clear: it  increases as the Universe cools down until a certain point where it freezes in,  remaining constant afterwards. As illustrated by the figure, the freeze in point depends on the dark matter mass. It is also clear that the rate at which $Y$ increases depends on both  $\msinglet$ and the temperature. At high temperature the singlet behaves as a relativistic particle so  $Y$ is pretty much independent of $\msinglet$, as observed in the figure for $T>1~\tev$. For $\msinglet=100~\gev$ and $\msinglet=1~\tev$, the dark matter abundance freezes in at $T\sim\msinglet$, when the particles in the plasma no longer have enough energy to produce the dark matter particles. For $\msinglet=10~\gev$ and $\msinglet=1~\gev$ the effect of the higgs resonance, which increases $\sv$ significantly, becomes important. That is why $Y$ increases steeply for $T\sim100~\gev$ before freezing in at  about $30~\gev$.  The reason  the abundance does not keep increasing until $T\sim \msinglet$ for $\msinglet=1~\gev$ is that, as stated before, the production of singlet dark matter is dominated by the $W^+W^-$ final state, which ceases to be abundant in the thermal plasma well before $T\sim 1~\gev$. For the same reason, the abundances are exactly the same for $\msinglet=1~\gev$ and $\msinglet=10~\gev$.

\begin{figure}[t]
\begin{center}
\includegraphics[scale=0.42]{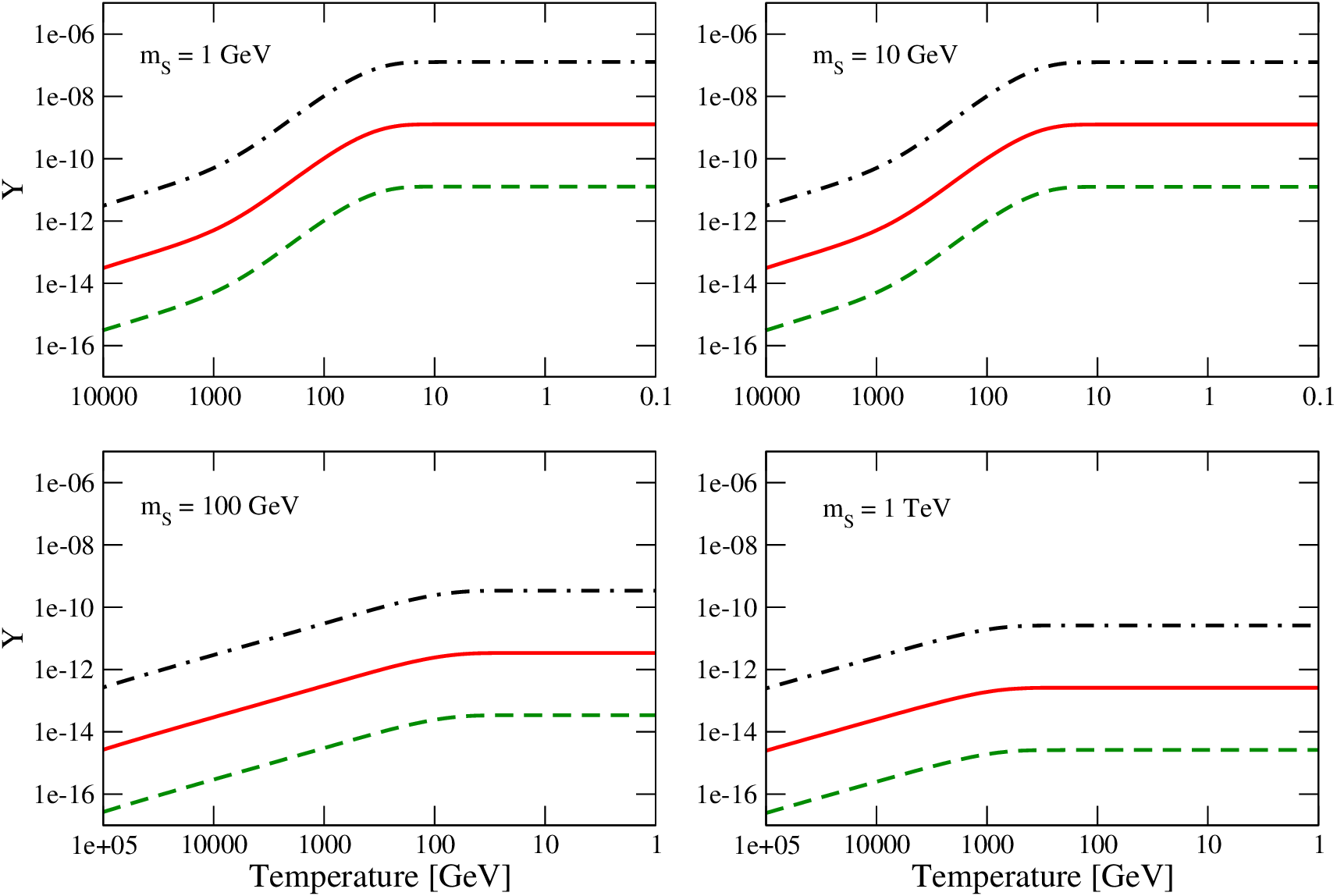}
\caption{\small \it The dark matter abundance, $Y$, as a function of the temperature for different values of $\lambda$ and of $\msinglet$. From top to bottom on each panel, the lines correspond to $\lambda=10^{-10}$ (dash-dotted), $\lambda=10^{-11}$ (solid), and $\lambda=10^{-12}$ (dashed). The higgs mass was set to $120~\gev$.
\label{fig:yarray}}
\end{center}
\end{figure}
\begin{figure}[t]
\begin{center}
\includegraphics[scale=0.42]{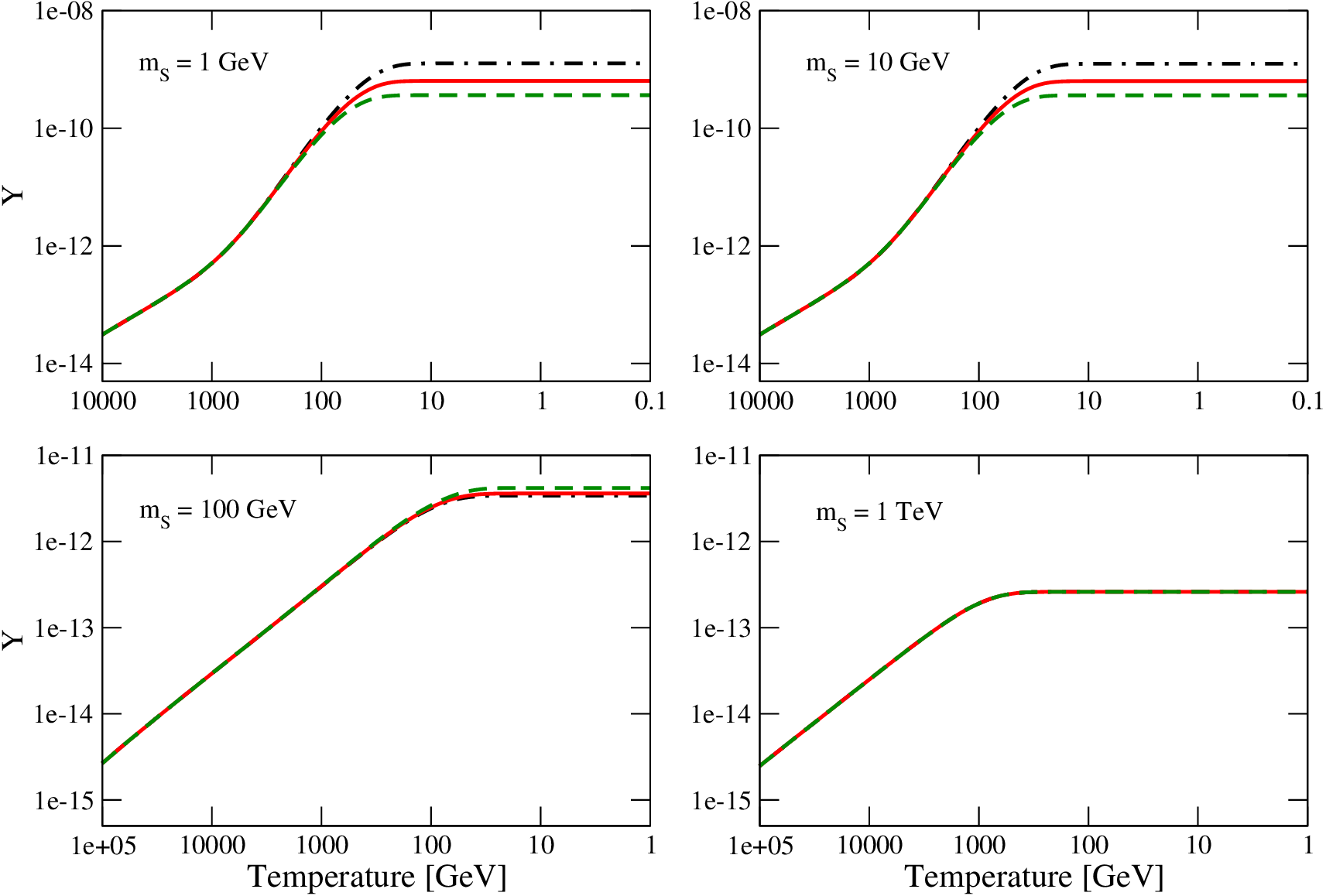}
\caption{\small \it The dark matter abundance,
 $Y$, as a function of the temperature for different values of $\mhiggs$ and of $\msinglet$. 
From top to bottom in each panel, the lines correspond to $\mhiggs=120~\gev$ (dash-dotted),  $\mhiggs=150~\gev$ (solid), and  $\mhiggs=180~\gev$ (dashed). In this figure $\lambda$ was chosen to be $10^{-11}$.
\label{fig:ymharray}
}
\end{center}
\end{figure}

Let us now see how these results are modified when we vary $\lambda$ and $\mhiggs$. Figure \ref{fig:yarray} shows the dark matter abundance as a function of the temperature for four different values of $\msinglet$ ($1,10,100,1000~\gev$, one in each panel) and three values of $\lambda$: $10^{-10}$ (upper, dashed-dotted line), $10^{-11}$ (middle, solid line) and $10^{-12}$ (lower, dashed line). The new feature observed in this figure is that the dark matter abundance is, as expected, exactly proportional to $\lambda^2$. This is a distinctive feature of the FIMP regime: the dark matter yield increases with the interaction strength between the dark matter particle and the thermal plasma. We will see later that, within this regime of the model,  the correct relic density of dark matter can be obtained for $\lambda$ between $10^{-11}$ and $10^{-12}$.

The dependence of the dark matter abundance with the higgs mass is illustrated in figure \ref{fig:ymharray}. It displays $Y$ as a function of the temperature for four different values of $\msinglet$ ($1,10,100,1000~\gev$, one in each panel) and three values of the higgs mass: $120~\gev$ (upper, dashed-dotted line), $150~\gev$ (middle, solid line) and $180~\gev$ (lower, dashed line). We see that for heavy singlets, which freeze in at $T\gg\mhiggs$, the abundance does not depend on the higgs mass at all. For lighter particles, $\msinglet=1,10~\gev$, the effect of different higgs masses become manifest at low temperatures $T\lesssim 100~\gev$. The smaller the higgs mass, the larger the freeze in temperature and the final dark matter abundance, as observed in the figure. In any case, since the higgs mass cannot vary over a wide range, the corresponding dark matter abundance does not change much either with $\mhiggs$. Its total effect amounts to a variation of less than one order of magnitude in $Y$.

In the FIMP regime, as the previous figures illustrate, the dark matter particles are slowly produced as the Universe cools down until the freeze in temperature is reached. In the singlet model, the freeze in temperature is given by the singlet mass for $\msinglet\gtrsim 100~\gev$ and by $T\sim 30\mbox{-}50~\gev$ for lighter singlets. Close to the higgs resonance the production accelerates but the abundance remains small ($Y\ll 1$) throughout the entire history of the Universe. In the FIMP regime, the dark matter particles never reach thermal equilibrium  or  annihilate among themselves. 

\begin{figure}[tb]
\begin{center}
\includegraphics[scale=0.42]{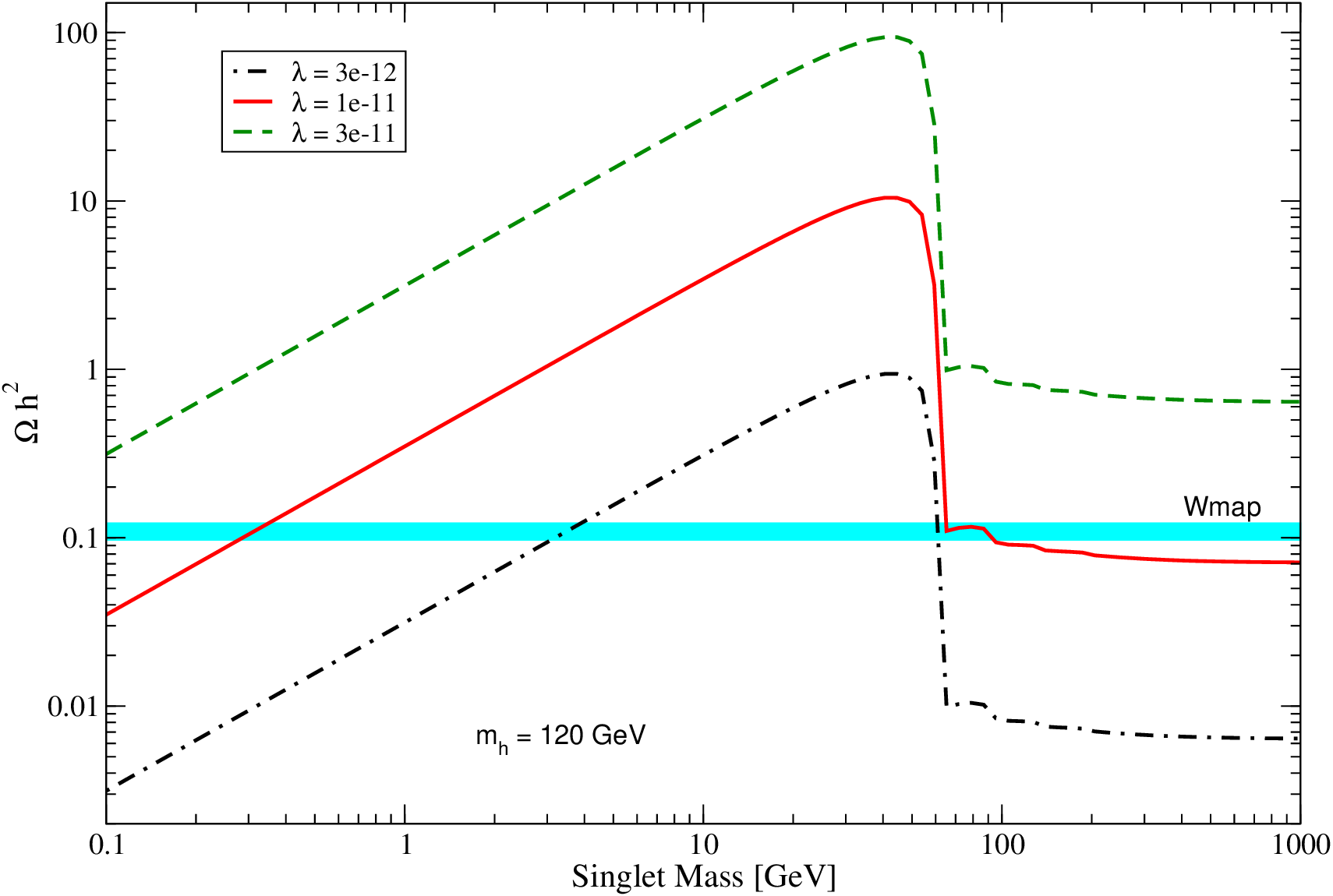}
\caption{\small \it The relic density of dark matter, $\oh$, as a function of $\msinglet$ for different values of $\lambda$. The higgs mass, $\mhiggs$, was fixed to $120~\gev$ in this figure. The Wmap compatible region is also shown as a horizontal band.
\label{fig:rd}}
\end{center}
\end{figure}

Now that we have analyzed the evolution of the dark matter abundance in the FIMP regime, we turn our attention to the relic density, the quantity that is actually constrained by observational data. The present relic density of dark matter is proportional to $\msinglet$ and to the asymptotic value of $Y$, see equation (\ref{eq:rd}). Figure \ref{fig:rd} displays the predicted relic density as a function of the singlet mass for $\mhiggs=120~\gev$ and different values of the coupling $\lambda$. From top to bottom the lines correspond to $\lambda=3\times 10^{-11}$ (dashed line), $\lambda=10^{-11}$ (solid line), and $\lambda=3\times 10^{-12}$ (dashed-dotted line). For comparison, the region compatible with the Wmap determination of the dark matter density \cite{Komatsu:2010fb} is also shown as a  horizontal band. As anticipated, the relic density is proportional to $\lambda^2$. Notice that for $\lambda=3\times 10^{-11}$ (the dashed line) the relic density is, independently of $\msinglet$, too high to be compatible with the Wmap range . For the other two values of $\lambda$ shown, we do find regions consistent with the data. Typically, in the singlet model there are two different masses that are consistent with the dark matter constraint for a given value of $\lambda$. It is clearly observed in the figure that the relic density has a sizable jump at the higgs resonance,  $\msinglet\sim 60~\gev$, or more generally at $\msinglet\sim \mhiggs/2$. If $\msinglet$ is slightly below the resonance the relic density is significantly larger than if it is slightly above the resonance. Moreover, while  above the resonance the relic density is essentially independent of the singlet mass, it becomes  proportional to $\msinglet$ below the higgs resonance. That is exactly the region where we have found the abundance, $Y$, to be independent of $\msinglet$ --see previous figures.    

\begin{figure}[tb]
\begin{center}
\includegraphics[scale=0.42]{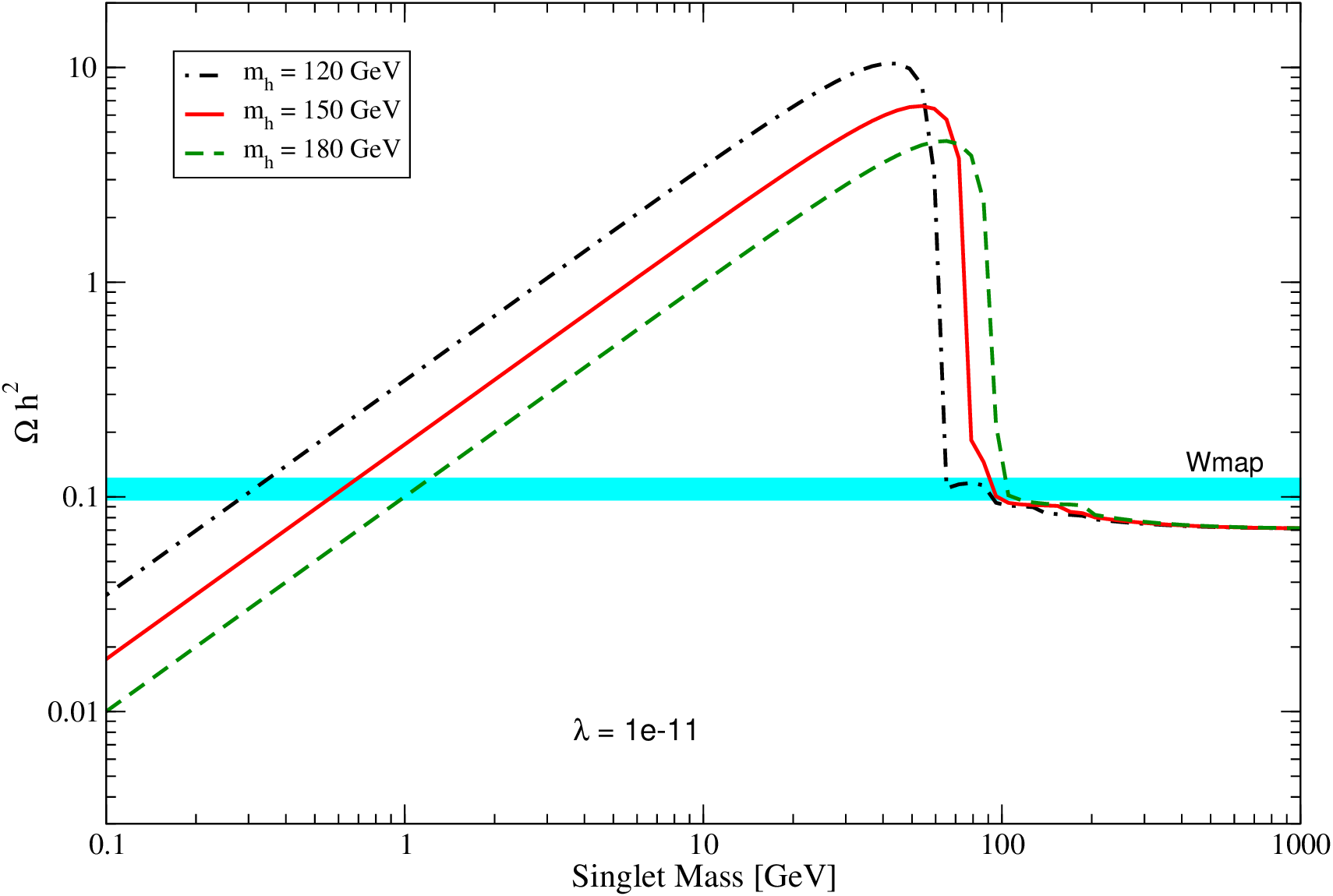}
\caption{\small \it The relic density of dark matter, $\oh$, as a function of $\msinglet$ for different values of $\mhiggs$. In this figure the coupling $\lambda$ was set to $10^{-11}$. The Wmap compatible region is also shown as a horizontal band.
\label{fig:rdmh}}
\end{center}
\end{figure}

The variation of the relic density with the higgs mass is shown in figure \ref{fig:rdmh}. We can see that the region where the behavior of the relic density changes is indeed determined by the higgs resonance condition: $\msinglet\sim \mhiggs/2$. For high masses, $\msinglet>\mhiggs$, the relic density is seen to be independent of $\mhiggs$. For lower masses, $\msinglet<\mhiggs/2$, the relic density turns out to be about a factor $3$ larger for $\mhiggs=120~\gev$ than for $\mhiggs=180~\gev$. Hence, in that region, the value of the mass that is compatible with the relic density might change by the same factor as $\mhiggs$ varies.

\begin{figure}[tb]
\begin{center}
\includegraphics[scale=0.42]{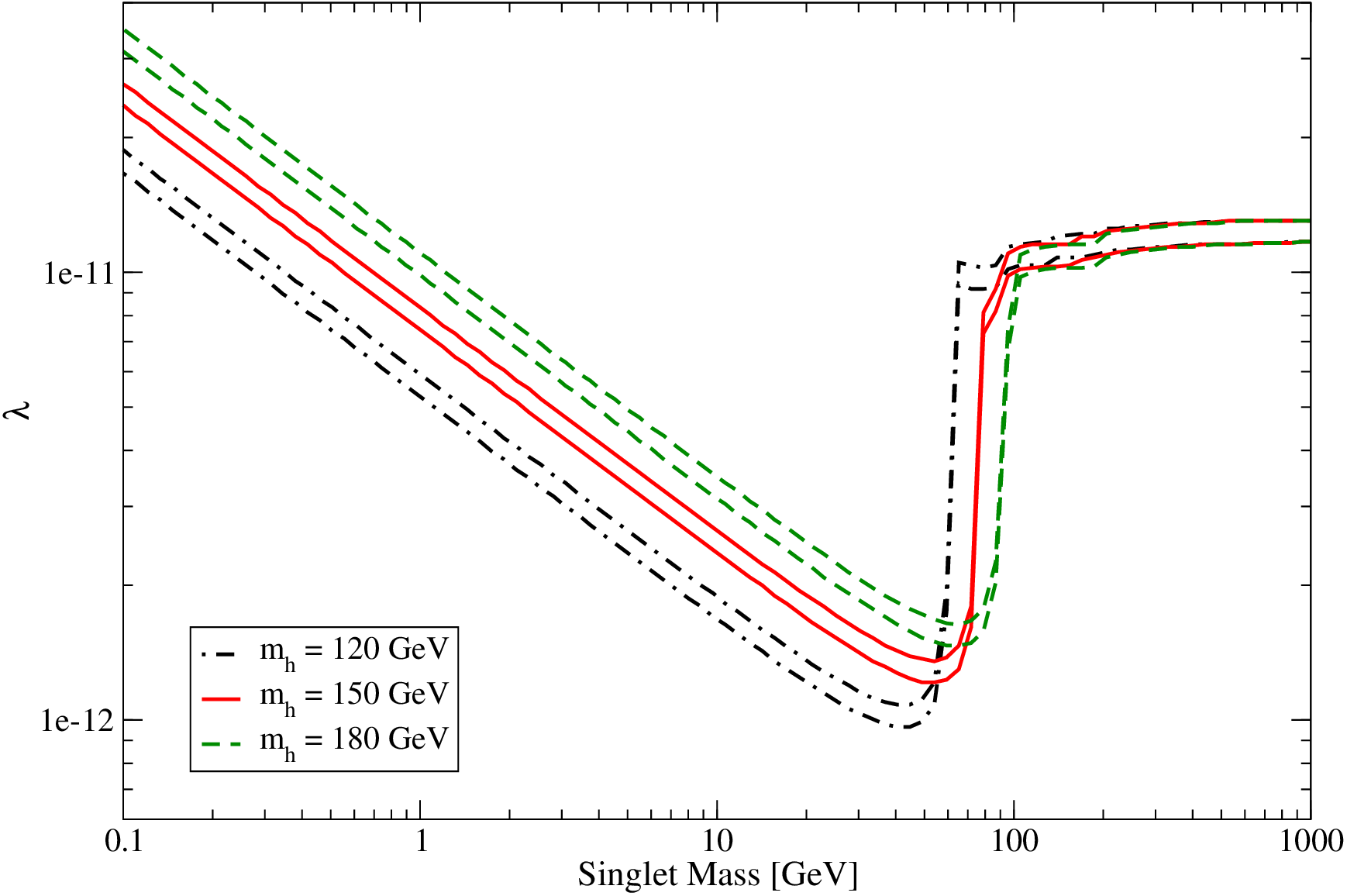}
\caption{\small \it The viable parameter space of the singlet model in the FIMP regime. Inside the bands the dark matter constraint is satisfied for the given value of the higgs mass. 
\label{fig:levelalt}}
\end{center}
\end{figure}

Usually we are interested not in computing the relic density by itself but rather in obtaining the viable parameter space of the model, the regions where the predicted relic density is consistent with the observed dark matter density. Figure \ref{fig:levelalt} shows, in the plane ($\msinglet,\lambda$), the viable region of the singlet model for the FIMP regime (see e.g. \cite{Yaguna:2008hd} for the corresponding regions in the WIMP regime). Each band shows the region compatible with the dark matter constraint (at $2\sigma$) for a given higgs mass. Notice that for heavy dark matter, $\msinglet>100~\gev$, $\lambda$ should be about $1.2\times 10^{-11}$ independently of the singlet mass or the higgs mass. For dark matter masses between $0.1~\gev$ and $50~\gev$ the required $\lambda$ decreases from about $2\times10^{-11}$ to about $10^{-12}$, reaching this minimum value for $\msinglet\sim 40~\gev$ and $\mhiggs=120~\gev$. Around the higgs resonance, the value of $\lambda$ that satisfies the dark matter constraint varies by about one order of magnitude in a narrow range of $\msinglet$. Notice if $\msinglet\lesssim 100~\gev$ for any given value of $\lambda$ there usually exists two solutions for $\msinglet$. If, for instance, $\mhiggs$ is found to be $150~\gev$, for $\lambda=3\times 10^{-12}$ the dark matter mass could be either about $8~\gev$ or about $75~\gev$. In any case, the crucial point is that such solutions do exist.

This last figure clearly demonstrates our main result, that in the singlet model the dark matter constraint can also be satisfied in a previously unexplored region of the parameter space. In this new viable region we have unveiled, the interactions between the  singlet and the Standard Model fields are very suppressed and the singlet  behaves as a Feebly Interacting Massive Particle (FIMP) rather than as a WIMP.  We would like to emphasize that this FIMP  regime of the singlet model is as natural and predictive as the WIMP regime. There is no need to include new particles, to introduce non-thermal production mechanisms, or to modify the cosmological model in order to satisfy the dark matter constraint. From the dark matter point of view, therefore,  both regimes should be treated on equal footing.     
\section{Discussion}
A generic property of the new viable region we have found is  the absence of any direct or indirect detection signature of dark matter, a direct consequence of the feeble interactions of the dark matter particles.  This lack of experimental signatures, which may currently seem as a drawback, could become a desirable feature if in the next years dark matter is not detected. Such an event would force us to abandon the WIMP paradigm and to replace it with an alternative framework. Among those that have been considered so far, FIMP dark matter looks particularly attractive due to its simplicity and predictivity.  

Even though the FIMP regime of the singlet model cannot be directly probed in dark matter experiments, it can certainly be falsified. As soon as a credible signal of dark matter is observed in direct or indirect detection experiments, the FIMP regime of the singlet scalar model can be excluded as the explanation of dark matter.

Recently, it was found that the latest constraints on dark matter direct detection  by the XENON100 experiment \cite{Aprile:2011hi} exclude a  significant fraction of the parameter space of the singlet model --see figure 3 in \cite{Farina:2011bh}. In particular, $\msinglet<80~\gev$ seems to be strongly disfavored. We would like to stress, however, that such a bound applies only to the WIMP regime of the singlet model and that no comparable bound exists in the FIMP regime. In it, the region $\msinglet<80~\gev$, as well as the entire $\gev$ to $\tev$ mass range, is perfectly compatible with the dark matter constraint and with the absence of a dark matter signal in direct and indirect dark matter experiments.

Finally, we would like to mention that the analysis we have presented for the singlet scalar model might be  extended to other  models of dark matter. As it is clear from our previous discussion, the main requirement on such models is that the interactions of the dark matter particle depend on a free parameter. If that is the case, it is likely that in such models also exist a new dark matter compatible region where the relic density constraint is satisfied via freeze in of Feebly Interacting Massive Particles. 

\section{Conclusions}
The singlet scalar model is one of the simplest extensions of the Standard Model that can explain the dark matter. It contains just one additional field, which is neutral under the SM gauge group and odd under a new $Z_2$ symmetry, and it introduces only two free parameters: the dark matter mass and the coupling between the dark matter and the higgs, $\lambda$. We have demonstrated that in this model there exists a previously unexplored region of the parameter space where the dark matter constraint can naturally be satisfied.  In it, the dark matter behaves  as a Feebly Interacting Massive Particle, FIMP, and its relic density is the result of a thermal \emph{freeze in} in the early Universe.  We have examined in detail the FIMP regime of  the singlet scalar model. After reviewing the calculation of the relic density, the evolution of the dark matter abundance with the temperature was analyzed as a function of the parameters of the model. Then, we computed the dark matter relic density and obtained the new viable parameter space of this model. It features couplings of order $10^{-11}$-$10^{-12}$ for dark matter masses in the $\gev$ to $\tev$ range. Finally, we discussed the implications of this new viable region for the correct interpretation of recent experimental bounds.   
\section*{Acknowledgments}
I would like to thank Yann Mambrini for useful discussions. 
This work has been supported by DFG grant no.
WI 2639/2-1.

\bibliographystyle{unsrt}
\bibliography{darkmatter}

\begin{thebibliography}{10}

\bibitem{Griest:2000kj}
K.~Griest and M.~Kamionkowski.
\newblock {Supersymmetric dark matter}.
\newblock {\em Phys. Rept.}, 333:167--182, 2000.

\bibitem{Hooper:2007qk}
Dan Hooper and Stefano Profumo.
\newblock {Dark matter and collider phenomenology of universal extra
  dimensions}.
\newblock {\em Phys. Rept.}, 453:29--115, 2007.

\bibitem{McDonald:1993ex}
John McDonald.
\newblock {Gauge singlet scalars as cold dark matter}.
\newblock {\em Phys.Rev.}, D50:3637--3649, 1994.

\bibitem{Burgess:2000yq}
C.P. Burgess, Maxim Pospelov, and Tonnis ter Veldhuis.
\newblock {The Minimal model of nonbaryonic dark matter: A Singlet scalar}.
\newblock {\em Nucl.Phys.}, B619:709--728, 2001.

\bibitem{Barbieri:2006dq}
Riccardo Barbieri, Lawrence~J. Hall, and Vyacheslav~S. Rychkov.
\newblock {Improved naturalness with a heavy Higgs: An alternative road to LHC
  physics}.
\newblock {\em Phys. Rev.}, D74:015007, 2006.

\bibitem{LopezHonorez:2006gr}
Laura Lopez~Honorez, Emmanuel Nezri, Josep~F. Oliver, and Michel H.~G. Tytgat.
\newblock {The inert doublet model: An archetype for dark matter}.
\newblock {\em JCAP}, 0702:028, 2007.

\bibitem{Honorez:2010re}
Laura Lopez~Honorez and Carlos~E. Yaguna.
\newblock {The inert doublet model of dark matter revisited}.
\newblock {\em JHEP}, 09:046, 2010.

\bibitem{Bertone:2010at}
Gianfranco Bertone.
\newblock {The moment of truth for WIMP Dark Matter}.
\newblock {\em Nature}, 468:389--393, 2010.

\bibitem{Choi:2005vq}
Ki-Young Choi and Leszek Roszkowski.
\newblock {E-WIMPs}.
\newblock {\em AIP Conf.Proc.}, 805:30--36, 2006.

\bibitem{Feng:2003xh}
Jonathan~L. Feng, Arvind Rajaraman, and Fumihiro Takayama.
\newblock {Superweakly interacting massive particles}.
\newblock {\em Phys.Rev.Lett.}, 91:011302, 2003.

\bibitem{Hall:2009bx}
Lawrence~J. Hall, Karsten Jedamzik, John March-Russell, and Stephen~M. West.
\newblock {Freeze-In Production of FIMP Dark Matter}.
\newblock {\em JHEP}, 1003:080, 2010.

\bibitem{Yaguna:2008hd}
Carlos~E. Yaguna.
\newblock {Gamma rays from the annihilation of singlet scalar dark matter}.
\newblock {\em JCAP}, 0903:003, 2009.

\bibitem{Goudelis:2009zz}
A.~Goudelis, Y.~Mambrini, and C.~Yaguna.
\newblock {Antimatter signals of singlet scalar dark matter}.
\newblock {\em JCAP}, 0912:008, 2009.

\bibitem{Belanger:2010gh}
G.~Belanger, F.~Boudjema, P.~Brun, A.~Pukhov, S.~Rosier-Lees, et~al.
\newblock {Indirect search for dark matter with micrOMEGAs2.4}.
\newblock {\em Comput.Phys.Commun.}, 182:842--856, 2011.

\bibitem{Komatsu:2010fb}
E.~Komatsu et~al.
\newblock {Seven-Year Wilkinson Microwave Anisotropy Probe (WMAP) Observations:
  Cosmological Interpretation}.
\newblock {\em Astrophys.J.Suppl.}, 192:18, 2011.

\bibitem{Aprile:2011hi}
E.~Aprile et~al.
\newblock {Dark Matter Results from 100 Live Days of XENON100 Data}.
\newblock {\em Phys.Rev.Lett.}, 2011.

\bibitem{Farina:2011bh}
Marco Farina, Mario Kadastik, Duccio Pappadopulo, Joosep Pata, Martti Raidal,
  et~al.
\newblock {Implications of XENON100 results for Dark Matter models and for the
  LHC}.
\newblock 2011.

\end{thebibliography}

\end{document}